\documentstyle [12pt]{article}
\def\fnote#1#2{\begingroup\def\thefootnote{#1}\footnote{#2}\addtocounter
{footnote}{-1}\endgroup}
\def\inbar{\vrule height1.5ex width.4pt depth0pt}
\def\IB{\relax{\rm I\kern-.18em B}}
\def\IC{\relax\,\hbox{$\inbar\kern-.3em{\rm C}$}}
\def\ID{\relax{\rm I\kern-.18em D}}
\def\IE{\relax{\rm I\kern-.18em E}}
\def\IF{\relax{\rm I\kern-.18em F}}
\def\IG{\relax\,\hbox{$\inbar\kern-.3em{\rm G}$}}
\def\IH{\relax{\rm I\kern-.18em H}}
\def\II{\relax{\rm I\kern-.18em I}}
\def\IK{\relax{\rm I\kern-.18em K}}
\def\IL{\relax{\rm I\kern-.18em L}}
\def\IM{\relax{\rm I\kern-.18em M}}
\def\IN{\relax{\rm I\kern-.18em N}}
\def\IO{\relax\,\hbox{$\inbar\kern-.3em{\rm O}$}}
\def\IP{\relax{\rm I\kern-.18em P}}
\def\IQ{\relax\,\hbox{$\inbar\kern-.3em{\rm Q}$}}
\def\IR{\relax{\rm I\kern-.18em R}}
\def\ZZ{\relax{\sf Z\kern-.4em Z}}
\def\fnote#1#2{\begingroup\def\thefootnote{#1}\footnote{#2}\addtocounter
{footnote}{-1}\endgroup}
\def\beq{\begin{equation}}
\def\eeq{\end{equation}}

\def\notin{\ \hbox{{$\in$}\kern-.51em\hbox{/}}}
\def\a{\alpha}   \def\b{\beta}   \def\g{\gamma}  \def\d{\delta}
\def\e{\epsilon}   \def\k{\kappa}  \def\l{\lambda}
  \def\om{\omega}   \def\si{\sigma}
\def\Si{\Sigma}  
   \def\cD{{\cal D}}
   
 \def\cK{{\cal K}}  \def\cM{{\cal M}}
 \def\cO{{\cal O}} \def\cP{{\cal P}} 
\def\cR{{\cal R}}  
\def\tcK{{\tilde \cK}} 
\def\tk{{\tilde \k}}

\begin{document}
\hfill HD--THEP--92--7
\vskip .01truein
\vskip 1truein
\centerline{\Large TOPOLOGICAL PHASES OF THE HETEROTIC STRING
            \fnote{\star}{This work was supported in part by the
                          National Science Foundation under Grant No.
                          PHY--89--04035.}}
\vskip .8truein
\centerline{\sc Jack Morse and Rolf Schimmrigk
            \fnote{\diamondsuit}{Bitnet address: q25@dhdurz1.bitnet}}
\vskip .5truein
\centerline{\it Institut f\"ur Theoretische Physik,
                Universit\"at Heidelberg}
\centerline{\it Philosophenweg 16, 6900 Heidelberg, FRG}
\vskip .05truein
\centerline{and}
\vskip .05truein
\centerline{\it Institute for Theoretical Physics,
                University of California}
\centerline{\it Santa Barbara, CA 93106, USA}

\baselineskip=21pt

\vskip 1.6truein
\centerline{\bf ABSTRACT}
\vskip .2truein

We analyze the phase structure of topological Calabi--Yau manifolds
defined on the moduli space of instantons. We show in this framework
that topological vacua describe new phases of the Heterotic String
theory in which the flat directions corresponding to complex
deformations are lifted.
We also briefly discuss the phase structure of non--K\"ahler manifolds.

\renewcommand\thepage{}
\vfill
\eject

\parskip .1truein
\parindent=20pt
\pagenumbering{arabic}
\noindent
{\sc 1. Introduction}

\noindent
It is believed that a topological $\si$--model coupled to 2D topological
gravity may represent a new phase of an underlying string theory [1].
We investigate the nature of this phase by computing the scaling
behaviour
of the complete set of operators associated to the cohomology groups
of a
general Calabi--Yau manifold. In the $\si$--model corresponding to a
string theory compactified on a Calabi--Yau manifold these fields
become
marginal giving rise to degenerate string vacua. We show that in the
topological phase of these theories the complex deformations and their
gravitational descendents are no longer
marginal at any critical level. Hence the corresponding degeneracy
of the vacua is lifted.
Thus a picture reminiscent of the Higgs phase transition emerges: in the
`high temperature' symmetric topological phase of string theory certain
components of the order parameter vanish. In the critical, `broken' phase
these components freeze out and acquire nonvanishing vacuum expectation
values.

Motivated by the issue of mirror symmetry we also briefly analyze the
phase structure of non--K\"ahler manifolds. It is shown that the
operators
corresponding to the third cohomology of such spaces are marginal at
every critical level.

\vskip .5truein
\noindent
{\sc 2. Topologicial Calabi--Yau Manifolds via Intersection Theory}

\noindent
The configuration space of string theory is the space of maps
$\cM \equiv \{\Phi: \Si_g \hookrightarrow M\}$ defining the embedding of
the worldsheet $\Si_g$ into the target manifold $M$. The idea of ref. [2]
is to define the topological phase of string theory via intersection
theory
on the moduli space $\cM$. To this end recall that associated to a form
$\om_p \in H^p(M)$ on an $n$--dimensional manifold $M$ is its Poincar\'e
dual $C_{\om_p} \in H_{n-p}(M)$. $\om_p$ has a delta function property
with respect to the cycle $C_{\om_p}$, it essentially vanishes outside
$C_{\om_p}$. Hence the 2D--operator that results from a pullback of the
$p$--form of the manifold vanishes unless the embedding map $\Phi$ maps
the
location of the operator into the Poincar\'e dual of the form. This
immediately leads to a dimension count that nonvanishing correlation
functions
\fnote{1}{ Meaning intersections of hypersurfaces of the moduli space
           defined by the constraint on the embedding maps just
           mentioned. }
have to satisfy: every insertion of an operator
introduces two real moduli but for a $p$--form there are also $p$
constraints
on the instanton maps. Hence one can associate a ``ghost number"
 $(2-p)$ to the $p$--form $\om_p$.

The operator ring on a Calabi--Yau vacuum is relatively simple; since
there
are no 1--forms the only nonvanishing Betti numbers are
$b_0,b_2,b_3,b_4,b_6$.
We denote by $\cP$ the puncture operator, associated to the connectivity
of
the manifold. $\cK_m, \tcK_m$ denote the K\"ahler forms and their duals,
$\cD_i$ denote the elements of the third cohomology and $\cR$ is the dual of
the puncture operator.
Since the dimension of the moduli space of instantons embedded in a
three complex dimensional Calabi--Yau manifold is actually zero it
follows from the ghost numbers of the operators that the only
nonvanishing threepoint functions are
\begin{eqnarray}
<\cP\cP\cR>&=&\eta_{\cP \cR}~=~1 \nonumber \\
<\cP\cK_m\tcK_n>&=& \eta_{mn} \nonumber \\
<\cP\cD_i\cD_j>&=&\eta_{ij} \nonumber \\
<\cK_p \cK_q \cK_r> &=& C_{pqr}.
\end{eqnarray}
Assuming that these correlation functions are associated to a path
integral
in which deformations along the operators are introduced one can
integrate these equations on the small phase space parametrized by
$t_{0,\cP},t_{0,\cK_i},t_{0,\tcK_i},t_{0,\cD_i},t_{0,\cR}$
\fnote{2}{We use the notation of ref. [3].}.
The result

\beq
\begin{array}{r l r l}
<\cP \cP>  & = t_{0,\cR}              & <\cP \cR>  & = t_{0,\cP} \\
<\cP \cK_m> & = \eta_{mn}t_{0,\tcK_n} & <\cP \tcK_m>& =
                                              \eta_{mn}t_{0,\cK_n}\\
<\cK_p \cK_q& = C_{qpr} t_{0,\cK_r}   & <\cK_m \tcK_n> & =
                                               \eta_{mn}t_{0,\cP}\\
<\cP \cD_i> & = \eta_{ij} t_{0,\cD_j} & <\cD_i \cD_j> & =
                                               \eta_{ij}t_{0,\cP}
\end{array}
\eeq

is corrected by instantons. The instantons (if isolated) will only
generate corrections to the correlation functions of the second
cohomology group
\beq
<\cK_p \cK_q \cK_r> = C_{pqr} + \sum_a E_{pqr}^a
\eeq
where the sum denotes the contributions of the various instantons
\beq
E^a_{pqr} = \int_a \cK_p \int_a \cK_q \int_a \cK_r.
\eeq
Assuming for simplicity that there is only one (1,1) form
one finds for the multipoint correlation functions of the (1,1) form in
this case
\beq
<\cK^n> = c\delta_{n,3}+\sum_a c_a^n
\eeq
and hence the twopoint function becomes
\beq
<\cK \cK> = c t_{0,\cK} + \sum_a c_a^2 e^{c_a t_{0,\cK}}.
\eeq
These twopoint functions
           above is
are not independent but satisfy the constitutive equations
\begin{eqnarray}
<\cK \tcK> &=&<\cP \cR> \nonumber \\
<\cD_i \cD_j> &=& \eta_{ij} <\cP \cR> \nonumber \\
<\cK \cK> &=& c<\cP \tcK>+\sum_a c_a^2 e^{c_a <\cP \tcK>}.
\end{eqnarray}

  From the last equation it follows that $<\cP \tcK> \sim 1$ and
$<\cK \cK> \sim 1$ and hence
\beq
 \g_{\tcK} = \g_{str} =2\g_{\cK}
\eeq
because
$<\cO_1 \cdots \cO_r>\sim x^{2-\g_{str} +\sum_{i=1}^r (\g_{\cO_i}-1)}$.
With $<\cK \tcK> \sim x^{-\g_{str} + \g_{\cK} +\g_{\tcK}}$
 the constitutive equations then lead to the remaining relations
\beq
\g_{\cR}  = 3\g_{\cK}=2\g_{\cD_i}.
\eeq

The upshot of these relations is that the scaling dimensions of all the
operators in the theory are determined in terms of the single quantity
$\g_{str}$:
\beq
\g({\cO_{\om_p}})=\frac{p}{4}\g_{str}.
\eeq

In addition, by use of the genus recursion relations
\beq
<\si_k(\cO_{\a}) \cO_{\b} \cO_{\g}> =
k<\si_{k-1}(\cO_{\a})\cO_{\d}>\eta^{\d \e} <\cO_{\e} \cO_{\b} \cO_{\g}>
\eeq
we can further determine the scaling dimensions of the gravitational
descendents $\si_n(\cO_{\a})$ in terms of $\g_{str}$
\beq
\g(\si_n(\cO_{\om_p})) =\frac{p+2n}{4}\g_{str}.
\eeq

The anomalous string dimension itself is determined by the choice of
the coupling constants $t_{k,\a}$ in the massive planar string equation
\beq
<\cP~\cO_{\a}> = \eta_{\a \b} t_{0,\b} +
               \sum_{k=1}^{\infty} \sum_{\b} k t_{k,\b}
               <\si_{k-1}(\cO_{\b})\cO_{\a}> \label{stringeq}.
\eeq
 This equation can be reduced, via the recursion relations,
to an equation purely in terms of the two point functions of the order
parameters of the theory.
Choosing the coupling constants as
\begin{eqnarray}
Directions~ of~ order~parameters &:&
t_{0,\cP} \equiv x,~~ t_{0,\cK} \equiv \k,~~t_{0,\tcK} \equiv \tk,~~
t_{0,\cD_i} \equiv \e_i,~~t_{0,\cR} \equiv \l \nonumber \\
Directions~ of~ deformations &:&
t_{1,\cP} \equiv 1,~~t_{k,\cP} \equiv -\frac{1}{k}~~,~
{\rm all~ others~ zero}
\end{eqnarray}
leads to the general planar string equations at level $k$ of a
$dim_{\IC}=3$ Calabi--Yau manifold with $b_2=1$
\fnote{3}{Terms with negative exponents at low $k$ should be set
          to zero.}
\begin{small}
\begin{eqnarray}
x  &=&  \frac{1}{k}<\cP \cR>^k \nonumber \\
\k &=& <\cP \tcK> <\cP \cR>^{k-1}\nonumber \\
\e_i&=& <\cP \cD_i> <\cP \cR>^{k-1}\nonumber \\
\tk&=& <\cP \cK><\cP \cR>^{k-1}
   + (k-1)<\cP \cR>^{k-2} \left\{\frac{c}{2}<\cP \tcK>^2
   + \sum_a c_a \right\}
                                                       \nonumber \\
  & &~~~~~~~~+ (k-1)<\cP \cR>^{k-2}
 \left\{\sum_a\left(c_a<\cP \tcK>-1\right)c_a e^{c_a<\cP \tcK>}\right\}
                                                        \nonumber \\
\l &=& <\cP \cP><\cP \cR>^{k-1}
       + (k-1)<\cP \cR>^{k-2} \left\{ <\cP \cK><\cP \tcK>
         + \frac{1}{2}<\cP \cD_i>\eta^{ij}<\cD_j \cP> \right\}
                                                        \nonumber \\
 & &+(k-1)(k-2)<\cP \cR>^{k-3}\left\{\frac{c}{6}<\cP \tcK>^3
         + <\cP \tcK> \sum_a c_a
         + \sum_a\left(c_a<\cP \tcK>-2\right) e^{c_a<\cP \tcK>} \right\}
                                                         \nonumber \\
\end{eqnarray}
\end{small}
  From these equations one readily derives $<\cP\cP>\sim x^{-2/k}$ and
hence the anomalous string dimension at level $k$ is
\beq
\g_{str}=\frac{2}{k}.
\eeq
Combining this result with our previous result for the dimensions of
the gravitational descendents we find
\beq
\g_k(\si_n(\cO_{\om_p})) = \frac{p+2n}{2k}
\eeq
   from which we conclude that in the topological phase there are no
marginal operators associated to the complex deformation of the
Calabi--Yau manifold.

\vskip .5truein
\noindent
{\sc 3. Topological Phases on Non--K\"ahler Manifolds}

\noindent
In this section we analyze the topological behaviour of manifolds with
no K\"ahler deformations but with vanishing first Chern class. Because
of
the existence of Calabi--Yau manifolds with no complex deformations
these spaces are of obvious interest in the context of a possible
understanding of mirror symmetry in terms of intersection theory.
We assume that
the moduli space of instantons is again zero--dimensional and we wish
to investigate the nature of the critical points in such a framework.
Consider the operators $\cP, \cD_i, \cR$, corresponding to the puncture
operator, the third cohomology and the dual of the puncture operator.

Since there are no K\"ahler deformations it is obvious that the
behaviour
of these theories should resemble that of a 2--complex dimensional
Calabi--Yau manifold, the K3 surface. The constitutive equations
collapse to the one equation
\beq
<\cD_i \cD_j> =\eta_{ij} <\cP \cR>
\eeq
and the string equations simplify to
\begin{eqnarray}
x &=&\frac{1}{k}<\cP \cR>^k \nonumber \\
\e_i&=& <\cP \cD_i> <\cP \cR>^{k-1}\nonumber \\
\l &=&<\cP \cR>^{k-2}\left\{<\cP \cP><\cP \cR>
                +\frac{k-1}{2}<\cP \cD_i>\eta^{ij}<\cD_j \cP> \right\}.
\end{eqnarray}
   From this it follows that $\g_{\cR}=\g_{str}+\frac{1}{k}$,
$\g_{\cR}= 2 \g_{\cD_i}$ and
\beq
\g_{str} = 2 - \frac{1}{k}
\eeq
and hence the order parameters corresponding to the complex deformation
are
marginal at every level $k$. Thus it might be possible to establish an
explicit map between these theories and the ones discussed in the
previous section, if perhaps only for the marginal sector.

\vskip .5truein
\noindent
{\sc 4. Conclusion}

\noindent
We have shown that topological $\si-$models on Calabi--Yau manifolds
formulated via intersection theory on the moduli space of instantons
define
a new phase of string theory. In this new phase the space of order
parameters
of marginal operators collapses to a subspace. In the critical phase
the order parameters corresponding to the complex deformations acquire
nonvanishing vevs and the degeneracy of the various vacua is enlarged.
All of this of course is rather reminiscent of the Higgs phase
transitions
in which the order parameter acquires nonvanishing vevs as well.

Motivated by the issue of mirror symmetry in this topological framework
we
have also analyzed the phase structure of non--K\"ahler manifolds with
vanishing first Chern class. The anomalous dimensions of these theories
turn out to be different from the K\"ahler manifold and the
order parameters corresponding to the primary operators of the complex
deformation are always marginal for these theories.

\vskip .4truein
\noindent
{\sc Acknowledgements}

We thank R.Dijkgraaf and E.Witten for discussions and correspondence.

\vskip .4truein
\noindent
{\sc References}
\vskip -.2truein
\begin{enumerate}
\item E.Witten, Comm.Math.Phys. {\bf 118}(1988)411
\item E.Witten, Nucl.Phys. {\bf B340}(1990)281
\item R.Dijkgraaf and E.Witten, Nucl.Phys. {\bf B342}(1990)486
\end{enumerate}
\end{document}